\documentclass[a4paper,11pt]{article}

\usepackage[latin1]{inputenc}

\usepackage[T1]{fontenc}
\usepackage[dvips]{graphicx}
\usepackage{times}

\usepackage{pstricks}
\usepackage{calrsfs}
\usepackage{subeqn}

\usepackage{pifont}
\usepackage{oldgerm}
\usepackage{color} 
\usepackage{amssymb,amsbsy}
\usepackage{epic,eepic}
\usepackage{texdraw}
\usepackage{latexsym}

\usepackage{accents}

\newcommand{\nn}{\nonumber}

\begin{document}
\title{Lattice Schwarzian Boussinesq equation and two-component systems}
\author{Pavlos Xenitidis and Frank Nijhoff \\ 
School of Mathematics, University of Leeds, LS2 9JT, Leeds, UK \\
P.Xenitidis@leeds.ac.uk, F.W.Nijhoff@leeds.ac.uk}
\date{}
\maketitle

\begin{abstract}
Various new two-component systems related to the lattice Schwarzian Boussinesq equation are constructed in a systematic way from conservation laws. Their multidimensional consistency is demonstrated, Lax pairs, symmetries and conservation laws are derived and an auto-B{\"a}cklund transformation is constructed. Finally, Yang-Baxter maps from these systems are constructed.
\end{abstract}

\section{Introduction}

The main concern of this paper is the systematic derivation of two-component systems related to the lattice Schwarzian Boussinesq equation
which is the following 9-point scalar partial difference equation:
\begin{eqnarray}
S[g_{n,m}] &:=& \frac{(\Delta_n g_{n+1,m+2}) (\Delta_m g_{n,m+1}-\Delta_n g_{n,m+1}) (\Delta_m g_{n,m})}{(\Delta_m g_{n+2,m+1}) (\Delta_n g_{n+1,m}-\Delta_m g_{n+1,m}) (\Delta_n g_{n,m})}  \nn \\
&&\,\,\, - \,\frac{\alpha^3 (\Delta_n g_{n,m+2}) (\Delta_n g_{n,m+1})-\beta^3 (\Delta_m g_{n+1,m+1}) (\Delta_m g_{n,m+1})}{\beta^3 (\Delta_m g_{n+2,m}) (\Delta_m g_{n+1,m})-\alpha^3 (\Delta_n g_{n+1,m+1}) (\Delta_n g_{n+1,m})} = 0, \label{eq:sbsq}
\end{eqnarray}
first given in \cite{N1}. This equation can be written as a three-component system, cf. \cite{H,W}, however, it seems more natural to write the equation as a two-component system on the quadrilateral. A first example of such a two-component version of the Schwarzian Boussinesq equation was given in \cite{A}, based on a B{\"a}cklund transformation, which is
equivalent to the following system:
\begin{subequations}\label{sbsq-sys-0}
\begin{eqnarray} 
g_{n+1,m+1} &=& \frac{g_{n+1,m} \Delta_n f_{n,m} - g_{n,m+1} \Delta_m f_{n,m}}{f_{n+1,m}-f_{n,m+1}}\,,\\
f_{n+1,m+1} &=& \frac{\alpha^3 (\Delta_n f_{n,m}) (\Delta_n g_{n,m}) \,f_{n,m+1}-\beta^3 (\Delta_m f_{n,m}) (\Delta_m g_{n,m})\, f_{n+1,m}}{\alpha^3 (\Delta_n f_{n,m}) (\Delta_n g_{n,m})-\beta^3 (\Delta_m f_{n,m}) (\Delta_m g_{n,m})}\,.
\end{eqnarray} 
\end{subequations}
In the present paper we shall give an alternative derivation of (\ref{sbsq-sys-0}) based on conservation laws. Furthermore, we shall show that there exist several alternative two-component systems related to the scalar lattice equation (\ref{eq:sbsq}), which we claim give a clear insight in the integrability properties.

The scalar Schwarzian Boussinesq equation, like any other higher-order system, can be written as a system of lower order equations in an arbitrary number of ways. However, there are only few canonical ways of writing integrable equations as systems in such a way that the system reflects the underlying structure of the equation. Here we present a systematic construction of systems which respect this structure.

Our derivation of these systems starts with another integrable system and relies on the conservation laws of this system and the interpretation of them as B{\"a}cklund transformation. More precisely, we use the integrable system related to the lattice modified Boussinesq system, see (\ref{mbsq-sys}) below, and the observation that both of the equations constituting this system can be written in the form of a conservation law. This allows us to introduce two potentials $g$ and $f$ and, subsequently, derive a system only for them. The advantage of this derivation is that it allows us to derive not only the system for the potentials but also a Lax pair, symmetries and canonical conservation laws for this system.

Furthermore, other systems can be derived from the Schwarzian Boussinesq system using the same idea of conservation laws and potentials. In particular, these system have the following forms
\begin{eqnarray*}
h_{n+1,m+1} &=& \frac{h_{n+1,m} f_{n,m+1} \Delta_n f_{n,m} - h_{n,m+1} f_{n+1,m} \Delta_m f_{n,m}}{f_{n,m} \left(f_{n+1,m}-f_{n,m+1}\right)}\,,\\
f_{n+1,m+1} &=& \frac{\alpha^3 (\Delta_n f_{n,m}) (\Delta_n h_{n,m}) \,f_{n,m+1}-\beta^3 (\Delta_m f_{n,m}) (\Delta_m h_{n,m})\, f_{n+1,m}}{\alpha^3 (\Delta_n f_{n,m}) (\Delta_n h_{n,m})-\beta^3 (\Delta_m f_{n,m}) (\Delta_m h_{n,m})}\,,
\end{eqnarray*}
and
\begin{eqnarray*}
&& g_{n+1,m+1} =  g_{n+1,m} g_{n,m+1} \frac{r_{n+1,m}-r_{n,m+1}}{g_{n,m+1} \Delta_n r_{n,m} - g_{n+1,m} \Delta_m r_{n,m}},\\
&& r_{n+1,m+1} = \frac{\alpha^3 g_{n,m+1} r_{n,m+1} \Delta_n g_{n,m} \Delta_n r_{n,m}-\beta^3 g_{n+1,m} r_{n+1,m} \Delta_m g_{n,m} \Delta_m r_{n,m}}{\alpha^3 g_{n,m+1} \Delta_n g_{n,m} \Delta_n r_{n,m}-\beta^3 g_{n+1,m} \Delta_m g_{n,m} \Delta_m r_{n,m}},
\end{eqnarray*}
respectively. They may be considered as another auto-B{\"a}cklund transformation of the Schwarzian Boussinesq equation since they can be decoupled for each of the fields involved in them to the same 9-point lattice scalar equation.

Another interesting outcome from the study of these systems is the derivation of Yang-Baxter maps along with corresponding Lax pairs. In this construction, we use as new variables the invariants of point symmetries of the system under consideration \cite{PTV}. The application of this approach to both of the above systems, as well as to their Lax pairs, results to some new Yang-Baxter maps along with their Lax matrices \cite{SV}.

The paper is organized as follows. In the next section we introduce our notation and give the definitions of a symmetry and a conservation law for a system of difference equations. Section \ref{sec-sbsq} contains the derivation of three lattice Schwarzian Boussinesq systems and their integrability properties. In Section \ref{sec-more} the derivation of an auto-B{\"a}cklund transformation for one of these systems is given. The last section deals with the construction of Yang-Baxter maps from the Boussinesq systems derived in Section \ref{sec-sbsq} along with corresponding Lax matrices.

\section{Notation}

In this section we introduce the notation we use throughout the paper and give some basic definitions in order to make our presentation self-contained.

We deal with systems of lattice equations involving two functions which depend on two independent discrete variables $n$, $m$. This dependence will be denoted by subscripts, for instance $u(n,m)$ is denoted by $u_{n,m}$ and $u(n+1,m)$ by $u_{n+1,m}$. The same notation will be used also for functions which depend implicitly on the independent variables, e.g. $A(u_{n,m},v_{n,m})$ will also be denoted simply by $A_{n,m}$. In any event, it will be clear from the text the dependence of every function on $n$ and $m$.

Let us denote by 
\begin{equation} \label{sys-gen}
\begin{array}{l}
   Q_1(u_{n,m},v_{n,m},u_{n+1,m},v_{n+1,m},u_{n,m+1},v_{n,m+1},u_{n+1,m+1}) = 0 \\
 Q_2(u_{n,m},v_{n,m},u_{n+1,m},v_{n+1,m},u_{n,m+1},v_{n,m+1},v_{n+1,m+1}) = 0 \end{array}
\end{equation}
a two-component system defined on an elementary quadrilateral of the lattice. A pair of functions $\rho_{n,m}$ and $\sigma_{n,m}$ which depend implicitly on $n$, $m$ is called a conservation law for system (\ref{sys-gen}) if
$$\Delta_m \rho_{n,m} = \Delta_m \sigma_{n,m}\qquad {\mbox{holds on solutions of system (\ref{sys-gen}).}}$$
Here, operators $\Delta_n$ and $\Delta_m$ are the difference operator in the $n$ and $m$ lattice direction, respectively, i.e. they act on a function $f_{n,m}$ as
$$\Delta_n f_{n,m} := f_{n+1,m}-f_{n,m},\quad \Delta_m f_{n,m} := f_{n,m+1}-f_{n,m}.$$ 
If the relation $\Delta_m \rho_{n,m} = \Delta_m \sigma_{n,m}$ holds identically, i.e. without taking into account equations (\ref{sys-gen}), we call this pair a trivial conservation law. For example, $\Delta_m \log(u_{n+1,m}/u_{n,m}) = \Delta_n \log(u_{n,m+1}/u_{n,m})$ is a trivial conservation law since it holds identically.

The pair of differential-difference equations
\begin{equation} \label{sym-gen}
\frac{{\rm{d}} u_{n,m}}{{\rm{d}} t}= \eta_{n,m},\quad \frac{{\rm{d}} v_{n,m}}{{\rm{d}} t}= \zeta_{n,m} 
\end{equation}
is a symmetry of system (\ref{sys-gen}) if they commute with the system. That is, if
$$\frac{{\rm{d}} Q_i}{{\rm{d}} t} = 0,\quad i=1,2, \qquad {\mbox{hold on solutions of system (\ref{sys-gen}).}}$$
In the last relation, the derivative of $Q_i$ is understood as a total derivative with respect to $t$, i.e.
$$\frac{{\rm{d}} Q_i}{{\rm{d}} t} := \sum_{k,l} \eta_{n+k,m+l} \frac{\partial Q_i}{\partial u_{n+k,m+l}} + \zeta_{n+k,m+l} \frac{\partial Q_i}{\partial v_{n+k,m+l}}. $$ 
Finally, let us recall the definition of the invariant of a symmetry: Any function $x_{n,m}$ which satisfies ${\rm{d}} x_{n,m}/{\rm{d}} t = 0 $ by taking into account equations (\ref{sym-gen}) is called an invariant of symmetry (\ref{sym-gen}).

\section{Two-component systems related to the Schwarzian Boussinesq equation} \label{sec-sbsq}

The conservation laws of a given integrable system can be used effectively in the construction of new integrable systems. The derivation of the continuous potential KdV equation could serve as the simplest and more illustrative example for this construction. In this section we apply this approach to the discrete case. In particular, starting with an integrable quadrilateral system, we derive a new one using a pair of conservation laws and demonstrate how the integrability aspects of the original system are mapped to corresponding characteristics of the new one. Here, the term integrability aspects mean Lax pair, symmetries and conservation laws.

The discrete system we start with is related to the lattice modified Boussinesq equation \cite{NPCQ} and it has the following form
\begin{subequations} \label{mbsq-sys}
\begin{eqnarray}
u_{n+1,m+1} &=& v_{n,m}\,\frac{\alpha u_{n,m+1} - \beta u_{n+1,m}}{\alpha v_{n+1,m} - \beta v_{n,m+1}}\,, \\
v_{n+1,m+1} &=& \frac{v_{n,m}}{u_{n,m}}\,\frac{\alpha u_{n+1,m} v_{n,m+1} - \beta u_{n,m+1} v_{n+1,m}}{\alpha v_{n+1,m} - \beta v_{n,m+1}}\,.
\end{eqnarray}
\end{subequations}
Its relation to the lattice modified Boussinesq equation is revealed when one decouples this system. More precisely, equations (\ref{mbsq-sys}) can be decoupled for $u_{n,m}$ leading to the 9-point equation 
\begin{eqnarray}
{\cal{M}}[u_{n,m}] &:=& \left(\frac{\alpha^2 u_{n+1,m+1} - \beta^2 u_{n,m+2}}{\alpha u_{n,m+2} - \beta u_{n+1,m+1}} \right) \frac{u_{n+1,m+2}}{u_{n,m+1}} - \left( \frac{\alpha^2 u_{n+2,m} - \beta^2 u_{n+1,m+1}}{\alpha u_{n+1,m+1} - \beta u_{n+2,m}} \right) \frac{u_{n+2,m+1}}{u_{n+1,m}}  \nn \\
\nn \\
&& - \alpha \,\left(\frac{u_{n,m}}{u_{n+1,m}} - \frac{u_{n+1,m+2}}{u_{n+2,m+2}}\right) -  \beta \,\left(\frac{u_{n,m}}{u_{n,m+1}} - \frac{u_{n+2,m+1}}{u_{n+2,m+2}}\right) = 0, \label{eq:mbsq}
\end{eqnarray}
which is known as the lattice modified Boussinesq equation \cite{NPCQ}. On the other hand, decoupling system (\ref{mbsq-sys}) for $v_{n,m}$, one finds 
that it obeys the equation
$${\cal{M}}\left[\frac{1}{v_{n,m}}\right] \,=\, 0\,.$$

Let us first note that both equations of system (\ref{mbsq-sys}) can be casted in the form of a conservation law. More precisely, the equations of system (\ref{mbsq-sys}) can be written as
\begin{subequations} \label{mbsq-sys-cf}
\begin{eqnarray}
&& \Delta_m\left( \frac{u_{n+1,m} v_{n,m}}{\alpha} \right)\,-\, \Delta_n\left( \frac{u_{n,m+1} v_{n,m}}{\beta} \right) = 0\,,\\
&& \Delta_m\left( \frac{u_{n,m}}{\alpha\,v_{n,m} v_{n+1,m}} \right)\,-\, \Delta_n\left( \frac{u_{n,m}}{\beta\,v_{n,m} v_{n,m+1}} \right)= 0.
\end{eqnarray}
\end{subequations}
The above relations allow us to introduce two potentials $g_{n,m}$ and $f_{n,m}$ through the relations
\begin{subequations} \label{pot-f-ps}
\begin{eqnarray}
&& \Delta_n g_{n,m} =  \frac{u_{n+1,m} v_{n,m}}{\alpha}\,, \quad  \Delta_m g_{n,m} =  \frac{u_{n,m+1} v_{n,m}}{\beta}\,,\label{pot-f-ps-1}\\
&& \Delta_n f_{n,m} =  \frac{u_{n,m}}{\alpha\,v_{n,m} v_{n+1,m}}\,, \quad \Delta_m f_{n,m} =  \frac{u_{n,m}}{\beta\,v_{n,m} v_{n,m+1}} \,.\label{pot-f-ps-2}
\end{eqnarray}
\end{subequations}

The compatibility conditions $\Delta_m(\Delta_n g_{n,m}) = \Delta_n (\Delta_m g_{n,m})$, $\Delta_m(\Delta_n f_{n,m}) = \Delta_n (\Delta_m f_{n,m})$ yield obviously system (\ref{mbsq-sys-cf}). On the other hand, if we solve equations (\ref{pot-f-ps}) for the shifted values of $u$ and $v$, i.e.
\begin{subequations} \label{pot-u-v}
\begin{eqnarray}
&& u_{n+1,m} =  \frac{\alpha \Delta_n g_{n,m} }{v_{n,m}}\,, \quad  u_{n,m+1} =  \frac{\beta \Delta_m g_{n,m} }{v_{n,m}}\,,\\
&& v_{n+1,m} =  \frac{u_{n,m}}{\alpha\,v_{n,m}  \Delta_n f_{n,m}}\,, \quad v_{n,m+1} =  \frac{u_{n,m}}{\beta\,v_{n,m} \Delta_m f_{n,m}} \,,
\end{eqnarray}
\end{subequations}
then the compatibility conditions ${\cal{S}}_m( u_{n+1,m} )= {\cal{S}}_n (u_{n,m+1})$, ${\cal{S}}_m (v_{n+1,m}) = {\cal{S}}_n (v_{n,m+1})$ lead to a coupled system for $g$ and $f$, which has the following form
\begin{subequations} \label{sbsq-sys}
\begin{eqnarray}
g_{n+1,m+1} &=& \frac{g_{n+1,m} \Delta_n f_{n,m} - g_{n,m+1} \Delta_m f_{n,m}}{f_{n+1,m}-f_{n,m+1}}\,,\\
f_{n+1,m+1} &=& \frac{\alpha^3 (\Delta_n f_{n,m}) (\Delta_n g_{n,m}) \,f_{n,m+1}-\beta^3 (\Delta_m f_{n,m}) (\Delta_m g_{n,m})\, f_{n+1,m}}{\alpha^3 (\Delta_n f_{n,m}) (\Delta_n g_{n,m})-\beta^3 (\Delta_m f_{n,m}) (\Delta_m g_{n,m})}\,.
\end{eqnarray}
\end{subequations}

This system is an involution of the lattice Schwarzian Boussinesq equation. That means, it relates solutions of the same equation since it can be decoupled either to the Schwarzian Boussinesq equation (\ref{eq:sbsq}) for $g$, i.e. $S[g_{n,m}] =0$, or to the same equation for $f$. The lattice Schwarzian Boussinesq equation was presented in \cite{N1}, its direct linearization derivation was developed in \cite{NPCQ}, where also the lattice modified Boussinesq equation appeared, and a three-component system related to it was given in \cite{W}. One-parameter extensions of these three-component systems were presented recently in \cite{H}.

System (\ref{sbsq-sys}) inherits its integrability aspects from system (\ref{mbsq-sys}); its Lax pair, its symmetries and its conservation laws follow by exploiting relations (\ref{pot-f-ps}) and (\ref{pot-u-v}). 

\begin{itemize}
\item Lax pair and multidimensional consistency

If we start with the Lax pair for system (\ref{mbsq-sys}), as it is given in \cite{XN}, make the gauge transformation 
$$G_{n,m} ={\rm{diag}}\left(v_{n,m},\frac{1}{u_{n,m}},\frac{u_{n,m}}{v_{n,m}}\right)$$ 
and then use relations (\ref{pot-u-v}), we end up with the following Lax pair for system (\ref{sbsq-sys}).
\begin{subequations}\label{lax-pair-sbsq-sys}
\begin{equation}
\Psi_{n+1,m} = L(\Delta_n g_{n,m},\Delta_n f_{n,m};\alpha) \Psi_{n,m} \,,\quad \Psi_{n,m+1} = L(\Delta_m g_{n,m},\Delta_m f_{n,m};\beta) \Psi_{n,m}\,,
\end{equation}
where
\begin{equation} \label{mbsq-L}
L(x,y;\alpha)  := 
\frac{1}{(\alpha^3-\lambda^3)^{1/3}} \left(\begin{array}{ccc} 
\alpha  & 0 &  -\lambda\,\alpha\,y\\
-\lambda \,\alpha\,x&\alpha & 0 \\
0 & \frac{-\lambda}{\alpha^2 x y} & \alpha
\end{array} \right)\,.
\end{equation}
\end{subequations}

Alternatively, one could employ the multidimensional consistency of system (\ref{sbsq-sys}) to derive the above Lax pair \cite{BS, N}. The multidimensional consistency of (\ref{sbsq-sys}) can be checked straightforwardly and it leads to the following relations for the triple-shifted values of $g$ and $f$ which are invariant under any permutation of the indices.
\begin{eqnarray*}
g_{123} &=& \frac{\alpha_1^3 F_1 G_1 \Big[g_2 F_2 - g_3 F_3 \Big] +\alpha_2^3 F_2 G_2 \Big[g_3 F_3 - g_1 F_1 \Big]  + \alpha_3^3 F_3 G_3 \Big[g_1 F_1 - g_2 F_2\Big]}{\alpha_1^3 F_1 G_1 (f_2 - f_3) + \alpha_2^3 F_2 G_2 (f_3 - f_1)+ \alpha_3^3 F_3 G_3 (f_1 - f_2)}\\
f_{123} &=& \frac{\alpha_1^3 \alpha_2^3 F_1 G_2 (g_1-g_2) f_3  + \alpha_2^3 \alpha_3^3 F_2 G_3 (g_2-g_3) f_1 + \alpha_3^3 \alpha_1^3 F_3 G_1 (g_3-g_1) f_2 }{\alpha_1^3 \alpha_2^3 F_1 G_2 (g_1-g_2)  + \alpha_2^3 \alpha_3^3 F_2 G_3 (g_2-g_3) + \alpha_3^3 \alpha_1^3 F_3 G_1 (g_3-g_1)}
\end{eqnarray*}
In the above formulas, we have used the following shorthand notations. Indices in $f$ and $g$ denote shifts in the corresponding lattice direction, e.g. $f_1 = f_{n+1,m,k}$ and $g_2 = g_{n,m+1,k}$. Functions $F_i$ and $G_i$ are defined by $F_i := f_i-f$ and $G_i:= g_i-g$, respectively. Finally, $\alpha_i$ denotes the lattice parameter corresponding to the $i$-lattice direction.

\item Symmetries and master symmetry

Similarly to the derivation of the Lax pair, we can start with the first generalized symmetry and the master symmetry of system (\ref{mbsq-sys}) \cite{XN} and use relations (\ref{pot-f-ps}) to derive corresponding symmetries for system (\ref{sbsq-sys}).

For instance, consider the first generalized symmetry
\begin{eqnarray*}
&& \frac{{\rm{d}} u_{n,m}}{{\rm{d}} t_1} =  \frac{3 u_{n,m} u_{n+1,m} v_{n,m}}{u_{n+1,m} v_{n,m} + u_{n,m} v_{n-1,m} + u_{n-1,m} v_{n+1,m}}-u_{n,m}\,,\\
&& \frac{{\rm{d}} v_{n,m}}{{\rm{d}} t_1} = \frac{-3 u_{n,m} v_{n-1,m} v_{n,m}}{u_{n+1,m} v_{n,m} + u_{n,m} v_{n-1,m} + u_{n-1,m} v_{n+1,m}}+v_{n,m}\,,
\end{eqnarray*}
of system (\ref{mbsq-sys}). If we differentiate relations (\ref{pot-f-ps}) with respect to $t_1$ and use the above relations to substitute the corresponding derivatives of $u$, $v$ and their shifts, a symmetry for the Schwarzian Boussinesq system will result. In the same fashion, a master symmetry follows from the master symmetry of system (\ref{mbsq-sys}). In particular, the resulting first generalized symmetry and master symmetry have the form
\begin{subequations} \label{sbsq-sym}
\begin{equation}
\frac{{\rm{d}} g_{n_,m}}{{\rm{d}} t_1}\,=\,(\Delta_n g_{n,m})\,A_{n,m},\quad \frac{{\rm{d}} f_{n,m}}{{\rm{d}} t_1}\,=\,(\Delta_n f_{n-1,m})\,A_{n,m}, 
\end{equation}
and
\begin{equation}
\alpha \frac{{\rm{d}} g_{n_,m}}{{\rm{d}} \alpha}\,=\,3\,n\,(\Delta_n g_{n,m})\,A_{n,m},\quad \alpha \frac{{\rm{d}} f_{n,m}}{{\rm{d}} \alpha}\,=\,3\,n\,(\Delta_n f_{n-1,m})\,A_{n,m}, 
\end{equation}
respectively, where
\begin{equation}
A_{n,m}\,:=\,\frac{\left(\Delta_n f_{n,m}\right) \left(\Delta_n g_{n-1,m}\right)}{\left(\Delta_n f_{n,m}\right) \left(\Delta_n g_{n,m}\right) + \left(\Delta_n f_{n,m}\right) \left(\Delta_n g_{n-1,m}\right) + \left(\Delta_n f_{n-1,m}\right) \left(\Delta_n g_{n-1,m}\right)}\,.
\end{equation}
In any case, it can be verified directly that the above equations define symmetries of system (\ref{sbsq-sys}).
\end{subequations}

A hierarchy of symmetries can be constructed successively by considering commutators. More precisely, the second symmetry follows from the commutator of the master symmetry and the first symmetry, while the third symmetry is the commutator of the second symmetry with the master symmetry. This procedure will result to the following hierarchy of symmetries
$$\frac{{\rm{d}} g_{n_,m}}{{\rm{d}} t_k}\,=\,{\cal{R}}^{k-1}\left\{(\Delta_n g_{n,m}) \,A_{n,m}\right\},\quad \frac{{\rm{d}} f_{n,m}}{{\rm{d}} t_k}\,=\,{\cal{R}}^{k-1}\left\{(\Delta_n f_{n-1,m}) \,A_{n,m}\right\}, \quad k=1,2,\cdots,$$
where
$${\cal{R}}:= \sum_{j=-\infty}^{\infty}j (\Delta_n g_{n+j,m})\,A_{n+j,m} \partial_{g_{n+j,m}} + \sum_{j=-\infty}^{\infty}j (\Delta_n f_{n-1+j,m})\,A_{n+j,m} \partial_{f_{n+j,m}} -\frac{\alpha}{3} \partial_\alpha.$$

\item Conservation laws

Finally, canonical conservation laws can be constructed in the same way. Specifically, if we add the trivial conservation law $\Delta_m \log(u_{n,m}/u_{n+1,m})^2 = \Delta_n\log(u_{n,m}/u_{n,m+1})^2$ to the conservation law $\Delta_m \rho_{n,m} = \Delta_n \sigma_{n,m}$ of system (\ref{mbsq-sys}), where
\begin{eqnarray*}
&& {\rho}_{n,m}\,=\,\log\left(\frac{u_{n+1,m} v_{n,m} + u_{n,m} v_{n-1,m} + u_{n-1,m} v_{n+1,m}}{u_{n,m} v_{n,m}}\right)^2\,,\\
&& {\sigma}_{n,m}\,=\,\log\left(\frac{\alpha^2 u_{n,m} v_{n-1,m} + \alpha \beta u_{n-1,m} v_{n,m+1} + \beta^2 u_{n,m+1} v_{n,m}}{(\alpha^3-\beta^3) u_{n,m} v_{n,m}}\right)^2\,,
\end{eqnarray*}
and then use relations (\ref{pot-u-v}), we will arrive at a conservation law for system (\ref{sbsq-sys}) with density $\varrho$ and flux $\chi$ given by
\begin{subequations} \label{sbsq-cl}
\begin{equation}
\varrho_{n,m} \,=\,\log \left(1 + \frac{(f_{n+1,m}-f_{n-1,m}) (g_{n,m}-g_{n-1,m})}{(f_{n+1,m}-f_{n,m}) (g_{n+1,m}-g_{n,m})} \right)^2
\end{equation}
and
\begin{equation}
\chi_{n,m} \,=\,\log \left(\frac{\beta^3}{\alpha^3-\beta^3} + \frac{\alpha^3 }{\alpha^3-\beta^3}\, \frac{(f_{n,m+1}-f_{n-1,m}) (g_{n,m}-g_{n-1,m})}{(f_{n,m+1}-f_{n,m}) (g_{n,m+1}-g_{n,m})} \right)^2,
\end{equation}
\end{subequations}
respectively. Moreover, higher order canonical conservation laws can be constructed recursively, starting with this conservation law and using the master symmetry, as it was done for the ABS equations \cite{X} and the Boussinesq systems \cite{XN}.
\end{itemize}

The covariance of the lattice Schwarzian Boussinesq system, i.e. its invariance under interchanges of the lattice directions, implies that another symmetry, master symmetry and a conservation law follow from the above ones by changing
$(n,m,u_{n+i,m+j},\alpha,\beta)$ to $(m,n,u_{n+j,m+i},\beta,\alpha)$. 

Now we can present two other systems related to the lattice Schwarzian Boussinesq equation. They are derived from system (\ref{sbsq-sys}) by introducing a new potential replacing either $g$ or $f$. This can be done by observing that the first equation of (\ref{sbsq-sys}) can be written also either as
$$\Delta_n\left( f_{n,m} \Delta_m g_{n,m}\right) \,=\, \Delta_m\left( f_{n,m} \Delta_n g_{n,m}\right),$$
or
$$\Delta_n\left( g_{n,m+1} \Delta_m f_{n,m}\right) \,=\, \Delta_m\left( g_{n+1,m} \Delta_n f_{n,m}\right). $$
If we use the first conservation law to introduce the function $h_{n,m}$ through the relations
\begin{equation} \label{rel-g-h}
\Delta_m h_{n,m} =  f_{n,m} \Delta_m g_{n,m},\quad \Delta_n h_{n,m} = f_{n,m} \Delta_n g_{n,m}\,,
\end{equation}
then we can eliminate $g$ and derive a system for $h_{n,m}$ and $f_{n,m}$, namely
\begin{subequations} \label{sbsq-sys-1}
\begin{eqnarray}
h_{n+1,m+1} &=& \frac{h_{n+1,m} f_{n,m+1} \Delta_n f_{n,m} - h_{n,m+1} f_{n+1,m} \Delta_m f_{n,m}}{f_{n,m} \left(f_{n+1,m}-f_{n,m+1}\right)},\\
f_{n+1,m+1} &=& \frac{\alpha^3 (\Delta_n f_{n,m}) (\Delta_n h_{n,m}) \,f_{n,m+1}-\beta^3 (\Delta_m f_{n,m}) (\Delta_m h_{n,m})\, f_{n+1,m}}{\alpha^3 (\Delta_n f_{n,m}) (\Delta_n h_{n,m})-\beta^3 (\Delta_m f_{n,m}) (\Delta_m h_{n,m})}.
\end{eqnarray}
\end{subequations}
Alternatively, we can use the second conservation law and introduce potential $r_{n,m}$ via
\begin{equation} \label{rel-f-r}
 \Delta_m r_{n,m} =  g_{n,m+1} \Delta_m f_{n,m}\,,\quad \Delta_n r_{n,m} = g_{n+1,m} \Delta_n f_{n,m}\,,
\end{equation}
and then eliminate $f$ to derive
\begin{subequations} \label{sbsq-sys-2}
\begin{eqnarray}
&& g_{n+1,m+1} =  g_{n+1,m} g_{n,m+1} \frac{r_{n+1,m}-r_{n,m+1}}{g_{n,m+1} \Delta_n r_{n,m} - g_{n+1,m} \Delta_m r_{n,m}},\\
&& r_{n+1,m+1} = \frac{\alpha^3 g_{n,m+1} r_{n,m+1} \Delta_n g_{n,m} \Delta_n r_{n,m}-\beta^3 g_{n+1,m} r_{n+1,m} \Delta_m g_{n,m} \Delta_m r_{n,m}}{\alpha^3 g_{n,m+1} \Delta_n g_{n,m} \Delta_n r_{n,m}-\beta^3 g_{n+1,m} \Delta_m g_{n,m} \Delta_m r_{n,m}}.
\end{eqnarray}
\end{subequations}
These multidimensionally consistent systems are related to the Schwarzian Boussinesq equation as well, i.e. they can be decoupled to the Schwarzian Boussinesq for either of the functions involved in them. Moreover, their integrability properties follow from the corresponding ones of system (\ref{sbsq-sys}) by employing relations (\ref{rel-g-h}) and (\ref{rel-f-r}), respectively. This can be done straightforwardly for the Lax pair (\ref{lax-pair-sbsq-sys}) and the canonical conservation law (\ref{sbsq-cl}). For symmetries, one has to use relations (\ref{rel-g-h}) and (\ref{rel-f-r}) along with (\ref{sbsq-sym}) in the same fashion as it is described above for the derivation of symmetries (\ref{sbsq-sym}).

Finally, let us mention some discrete symmetries of systems (\ref{sbsq-sys}), (\ref{sbsq-sys-1}) and (\ref{sbsq-sys-2}). It can be checked by straightforward calculations that system (\ref{sbsq-sys}) is invariant under the interchanges
$$\left(g_{n+i,m+j},f_{n+i,m+j} \right)\,\longrightarrow\,\left(f_{n-i,m-j},g_{n-i,m-j} \right)\,.$$
System (\ref{mbsq-sys}) is also invariant under a similar transformation and this property of these two systems will be explored further in the next section.
On the other hand, the other two systems are not invariant under such involutions but one is mapped to the other when we make similar interchanges. Indeed, if we make the changes
$$\left(h_{n+i,m+j},f_{n+i,m+j} \right)\,\longrightarrow\,\left(r_{n-i,m-j},g_{n-i,m-j} \right) $$
in system (\ref{sbsq-sys-1}) and then shift forward the resulting expressions in both lattice directions, we will derive system (\ref{sbsq-sys-2}). In the same fashion, the changes
$$\left(g_{n+i,m+j},r_{n+i,m+j} \right)\,\longrightarrow\,\left(f_{n-i,m-j},h_{n-i,m-j} \right) $$
map system (\ref{sbsq-sys-2}) to system (\ref{sbsq-sys-1}).

\section{Conservation laws and auto-B{\"a}cklund transformations} \label{sec-more}

So far, we have used two conservation laws of system (\ref{mbsq-sys}) to introduce two potentials and, subsequently, derive system (\ref{sbsq-sys}) for them. In other words, we have constructed from equations (\ref{mbsq-sys-cf}) a B{\"a}cklund transformation relating two systems. Here we employ the same idea in order to derive auto-B{\"a}cklund transformation for system (\ref{sbsq-sys}). More precisely, introducing a third potential, we construct a transformation which allows us to use one solution of system (\ref{sbsq-sys}) in order to find two new solutions of the same system.

We start by noting that system (\ref{mbsq-sys}) remains invariant under the interchanges
$$ (u_{n+i,m+j},v_{n+i,m+j})\,\longrightarrow\,(v_{n-i,m-j},u_{n-i,m-j})\,,\quad i,j=0,1\,.$$
This means that if we make these changes to system (\ref{mbsq-sys}) and then shift forward the resulting system in both lattice directions, we will arrive at the original system (\ref{mbsq-sys}). If we apply the same interchanges to (\ref{mbsq-sys-cf}), i.e. the conservation law form of system (\ref{mbsq-sys}), then the first equation remains invariant while the second one results to another conservation law, namely
\begin{equation}\label{sbsq-3-cl}
\Delta_m\left( \frac{v_{n+1,m}}{\alpha\,u_{n,m} u_{n+1,m}}\right) \,=\, \Delta_n\left( \frac{v_{n,m+1}}{\beta\,u_{n,m} u_{n,m+1}}\right)\,. 
\end{equation}
Now one can use any pair of the three conservation laws, i.e. the ones given in (\ref{mbsq-sys-cf}) and the above one, to derive a system for the corresponding potentials. 

Let us first rename potentials $g$ and $f$ defined in relation (\ref{pot-f-ps}) to $p^{(0)}$ and $p^{(1)}$, respectively, that is
\begin{subequations}
\begin{eqnarray}
&& \Delta_n p^{(0)}_{n,m} =  \frac{u_{n+1,m} v_{n,m}}{\alpha}\,, \quad  \Delta_m p^{(0)}=\frac{u_{n,m+1} v_{n,m}}{\beta}\,, \label{p0}\\
&& \Delta_n p^{(1)}_{n,m} =  \frac{u_{n,m}}{\alpha\,v_{n,m} v_{n+1,m}}\,, \quad \Delta_m p^{(1)}_{n,m} =  \frac{u_{n,m}}{\beta\,v_{n,m} v_{n,m+1}}. \label{p1}
\end{eqnarray}
Now we introduce a potential $p^{(2)}$ using conservation law (\ref{sbsq-3-cl}), i.e.
\begin{equation} \label{p2}
\Delta_np^{(2)}_{n,m}=\frac{v_{n+1,m}}{\alpha\,u_{n,m} u_{n+1,m}}\,, \quad  \Delta_m p^{(2)} = \frac{v_{n,m+1}}{\beta\,u_{n,m} u_{n,m+1}}.
\end{equation}
\end{subequations}
Then, using any pair of the relations (\ref{p0})--(\ref{p2}) to eliminate $u$ and $v$ from system (\ref{mbsq-sys}), we derive a system for the corresponding pair of potentials. In fact, the resulting system is
\begin{subequations}
\begin{eqnarray}
p^{(i)}_{n+1,m+1} &=& \frac{p^{(i)}_{n+1,m} \Delta_n p^{(i+1)}_{n,m} - p^{(i)}_{n,m+1} \Delta_m p^{(i+1)}_{n,m}}{p^{(i+1)}_{n+1,m}-p^{(i+1)}_{n,m+1}}\,,\\
p^{(i+1)}_{n+1,m+1} &=& \frac{\alpha^3 (\Delta_n p^{(i)}_{n,m}) (\Delta_n p^{(i+1)}_{n,m}) \,p^{(i+1)}_{n,m+1}-\beta^3 (\Delta_m p^{(i)}_{n,m}) (\Delta_m p^{(i+1)}_{n,m})\, p^{(i+1)}_{n+1,m}}{\alpha^3 (\Delta_n p^{(i)}_{n,m}) (\Delta_n p^{(i+1)}_{n,m})-\beta^3 (\Delta_m p^{(i)}_{n,m}) (\Delta_m p^{(i+1)}_{n,m})}\,,
\end{eqnarray}
\end{subequations}
where $i=0,1,2\,\,{\rm{mod}}\,\,3$. We will denote this system as $P\left(p^{(i)},p^{(i+1)}\right)$, or simply as $P_{i,i+1}$, and, apparently, one can identify $P_{0,1}$ with system (\ref{sbsq-sys}) and $P_{1,2}$ with the system derived in \cite{A}.

There is a simple B{\"a}cklund transformation among these systems which is a consequence of relations (\ref{p0})--(\ref{p2}) and reads as follows.
\begin{equation} \label{rotation}
{\mathbb{B}}\,:\,\left\{ \begin{array}{c} \alpha^3\,\Delta_n p^{(0)}_{n,m}\Delta_n p^{(1)}_{n,m}\Delta_n p^{(2)}_{n,m} \,=\,1 \\ \beta^3\,\Delta_m p^{(0)}_{n,m}\Delta_m p^{(1)}_{n,m}\Delta_m p^{(2)}_{n,m}\,=\,1 \end{array} \right.
\end{equation}
For every system $P$, the above relations define a B{\"a}cklund transformation mapping solutions of one system to solutions of another system. This means that if we start with any solution $(p^{(i)},p^{(i+1)})$ of $P_{i,i+1}$ and use relations (\ref{rotation}) to construct $p^{(i+2)}$, then the pairs $(p^{(i+1)},p^{(i+2)})$ and $(p^{(i+2)},p^{(i)})$ are solutions of $P_{i+1,i+2}$ and $P_{i+2,i}$, respectively. This is clear if we solve $\mathbb{B}$ for the differences of $p^{(i)}$ or $p^{(i+1)}$, respectively, and then use these expressions to eliminate the corresponding function from $P_{i,i+1}$ in favor of $p^{(i+2)}$. This elimination will result to $P_{i+1,i+2}$ or $P_{i+2,i}$, respectively. Since all systems $P$ have the same form, relations $\mathbb{B}$ define actually an auto-B{\"a}cklund transformation for these systems which differs from the natural one deriving from their multidimensional consistency.

\section{Symmetries and Yang-Baxter maps}

Multidimensional consistency is widely accepted as a notion of integrability, e.g. \cite{ABS, BS, N}, and it is very closely related to the Yang-Baxter property for maps \cite{ABS, PTV, SV}. In \cite{PTV}, this relation was explored systematically and it was shown how multidimensionally consistent systems and the invariants of their point symmetries can be used effectively in the derivation of Yang-Baxter maps. In this section we derive Yang-Baxter maps from systems (\ref{sbsq-sys}) and (\ref{sbsq-sys-1}) using the invariants of their points symmetries. Moreover, extending these symmetries to symmetries of the corresponding Lax pairs, we give Lax representations for the resulting Yang-Baxter maps. For a more detailed analysis, definitions and terminology on Yang-Baxter maps and Lax pairs one may refer to \cite{ABS-YB, Dr,PTV,SV,V}. 

Following \cite{PTV}, we employ the translation and scaling invariance of system (\ref{sbsq-sys}) to derive two Yang-Baxter maps, as well as the scaling invariance of (\ref{sbsq-sys-1}) to derive a third map. More precisely, in each case we introduce a pair of invariants on every edge of an elementary quadrilateral where the system is defined
$$x_i = x_i({\bf{f}}_{n,m},{\bf{f}}_{n+1,m}),\quad x_i^\prime = x_i({\bf{f}}_{n,m+1},{\bf{f}}_{n+1,m+1}), \quad i=1,2,$$
$$y_i = y_i({\bf{f}}_{n+1,m},{\bf{f}}_{n+1,m+1}),\quad y_i^\prime = y_i({\bf{f}}_{n,m},{\bf{f}}_{n,m+1}), \quad i=1,2,$$
where ${\bf{f}}_{n,m}$ stands for the pair of functions involved in the system. These invariants are not independent and  satisfy certain relations. For instance, among the invariants for the translation symmetry of system (\ref{sbsq-sys}) holds $x_i+y_i = x_i^\prime+y_i^\prime$, $i=1,2$. Taking into account the relations among the invariants and expressing the original quadrilateral system in terms of them, one can derive an invertible map $R : (x_i,y_i) \mapsto (x_i^\prime,y_i^\prime)$ and its companion ${\bar{R}} : (x_i,y_i^\prime) \mapsto (x_i^\prime,y_i)$. These maps are quadrirational \cite{ABS-YB}, have the reversibility property \cite{V} and satisfy the parameter-dependent Yang-Baxter equation
$$R_{23}(\beta,\gamma)R_{13}(\alpha,\gamma)R_{12}(\alpha,\beta)\, =\, R_{12}(\alpha,\beta) R_{13}(\alpha,\gamma) R_{23}(\beta,\gamma).$$

For the derivation of Lax pairs, we extend the points symmetries of these systems to symmetries of the corresponding Lax pairs and construct an invariant for the potential. To make this clear, let us consider the Lax pair (\ref{lax-pair-sbsq-sys}) of system (\ref{sbsq-sys}) and the scaling symmetry of the latter. First we extend this symmetry to a symmetry of the Lax pair
$$\frac{{\rm{d}} g_{n,m}}{{\rm{d}} t} = g_{n,m},\quad \frac{{\rm{d}} f_{n,m}}{{\rm{d}} t} = f_{n,m}, \quad \frac{{\rm{d}} \Phi_{n,m}}{{\rm{d}} t} = J_{n,m} \Phi_{n,m},$$
where matrix $J_{n,m}$ is determined by the equations
$$J_{n+1,m}L(\alpha) = \frac{{\rm{d}}}{{\rm{d}} t} L(\alpha) + L(\alpha) J_{n,m}, \quad J_{n,m+1}L(\beta) = \frac{{\rm{d}}}{{\rm{d}} t} L(\beta) + L(\beta) J_{n,m}\,.$$
Here we have omitted the dependence of Lax matrix $L$ of (\ref{lax-pair-sbsq-sys}) on $g$ and $f$ for simplicity. 

Then we solve these equations for matrix $J_{n,m}$ to find that $J_{n,m} = {\rm{diag}}(0,1,-1)$, and introduce the invariant $\Psi$ determined by
$$\Psi_{n,m} = H_{n,m} \Phi_{n,m}\,\quad {\mbox{where}} \quad g_{n,m} \frac{\partial H_{n,m}}{\partial g_{n,m}} + f_{n,m} \frac{\partial H_{n,m}}{\partial f_{n,m}} + H_{n,m} J_{n,m} = 0. $$
In particular, this equation for $H_{n,m}$ implies that
$$H_{n,m}\,=\,{\rm{diag}}\left(1, \frac{1}{g_{n,m}}, f_{n,m} \right)\,.$$
Plugging all the invariant forms back into the Lax pair, we arrive at
$$M(x,y;\alpha)\,:=\,H_{n+1,m} L(\alpha) H_{n,m}^{-1},$$
which is a Lax matrix for the corresponding Yang-Baxter map, i.e. relation
$$ M(x_1^\prime,x_2^\prime;\alpha) M(y_1^\prime,y_2^\prime;\beta) =  M(y_1,y_2;\beta)  M(x_1,x_2;\alpha) $$
implies the Yang-Baxter map \cite{SV}.

We finish this note by presenting a list of the various integrability characteristics that
have emerged, namely the group invariants we have used for the dependent variables, the Yang-Baxter maps and the relevant Lax matrices.

\begin{itemize}
\item Translation invariance of system (\ref{sbsq-sys})

\begin{enumerate}
\item Invariants
\begin{eqnarray*}
x_1 = \Delta_n f_{n,m},\quad x_2 = \Delta_n g_{n,m},\quad y_1 = \Delta_m f_{n+1,m}, \quad y_2 = \Delta_m g_{n+1,m}\,,\\
x_1^\prime = \Delta_n f_{n,m+1},\quad x_2^\prime = \Delta_n g_{n,m+1},\quad y_1^\prime = \Delta_m f_{n,m}, \quad y_2^\prime = \Delta_m g_{n,m}\,,
\end{eqnarray*}
\item Yang-Baxter map
\begin{eqnarray*}
&& x_1^\prime \,=\,\frac{\beta^3 y_1 (x_1 x_2 + y_1 y_2 + y_1 x_2)}{\alpha^3 x_1 x_2 + \beta^3 y_1 (x_2+y+2)}\,, \nn \\
&& x_2^\prime \,=\,\frac{y_2 (\alpha^3 x_1 x_2 + \beta^3 y_1 (x_2+y_2))}{\alpha^3 x_2 (x_1+y_1) + \beta^3 y_1 y_2}\\
&& y_1^\prime \,=\, x_1 + y_1 - x_1^\prime,\quad y_2^\prime \,=\,x_2+y_2-x_2^\prime.
\end{eqnarray*}
\item Lax matrix for the Yang-Baxter map
$$ M(x,y;\alpha) \,=\,\frac{1}{(\alpha^3-\lambda^3)^{1/3}} \left( \begin{array}{ccc} \alpha & 0 & -\alpha \lambda x \\ -\alpha \lambda y & \alpha & 0 \\ 0 & \frac{-\lambda}{\alpha^2 x y} & \alpha \end{array} \right) $$
\end{enumerate}

\item Scaling invariance of system (\ref{sbsq-sys})

\begin{enumerate}
\item Invariants
\begin{eqnarray*}
x_1 = \frac{f_{n+1,m}}{f_{n,m}},\quad x_2 = \frac{g_{n+1,m}}{g_{n,m}},\quad y_1 = \frac{f_{n+1,m+1}}{f_{n+1,m}}, \quad y_2 = \frac{g_{n+1,m+1}}{g_{n+1,m}}\,,\\
x_1^\prime = \frac{f_{n+1,m+1}}{f_{n,m+1}},\quad x_2^\prime = \frac{g_{n+1,m+1}}{g_{n,m+1}},\quad y_1^\prime = \frac{f_{n,m+1}}{f_{n,m}}, \quad y_2^\prime = \frac{g_{n,m+1}}{g_{n,m}}\,,
\end{eqnarray*}
\item Yang-Baxter map
$$ x_1^\prime\,=\,y_1\,A,\quad y_1^\prime\,=\,x_1\,\frac{1}{A},\quad x_2^\prime\,=\,y_2\,B,\quad y_2^\prime\,=\,x_2\,\frac{1}{B}, $$
where
\begin{eqnarray*}
&&  A:= \frac{\alpha^3 (x_1-1) (x_2-1) + \beta^3 x_1 (y_1-1) (x_2 y_2-1)}{\alpha^3 (x_1-1) (x_2-1) y_1 + \beta^3  (y_1-1) (x_2-1 + x_1 x_2 (y_2-1))} \,,\\
&& B := \frac{\alpha^3 (x_2-1) (x_1 y_1-1) + \beta^3 x_1 x_2 (y_1-1) (y_2-1)}{\alpha^3 (x_2-1) (x_1-1 + x_1 y_2 (y_1-1)) + \beta^3  x_1 (y_1-1) (y_2-1)} .
\end{eqnarray*}
\item Lax matrix for the Yang-Baxter map
$$  M(x,y;\alpha) \,=\,\frac{1}{(\alpha^3-\lambda^3)^{1/3}} 
\left( \begin{array}{ccc} \alpha & 0 & \alpha \lambda (1-x) \\ 
\frac{\alpha \lambda (1-y)}{y} & \frac{\alpha}{y} & 0 \\ 
0 & \frac{-\lambda x}{\alpha^2 (1-x) (1-y)} & \alpha x \end{array} \right)$$
\end{enumerate}

\item Scaling invariance of system (\ref{sbsq-sys-1})

\begin{enumerate}
\item Invariants
\begin{eqnarray*}
x_1 = \frac{f_{n+1,m}}{f_{n,m}},\quad x_2 = \frac{h_{n+1,m}}{h_{n,m}},\quad y_1 = \frac{f_{n+1,m+1}}{f_{n+1,m}}, \quad y_2 = \frac{h_{n+1,m+1}}{h_{n+1,m}}\,,\\
x_1^\prime = \frac{f_{n+1,m+1}}{f_{n,m+1}},\quad x_2^\prime = \frac{h_{n+1,m+1}}{h_{n,m+1}},\quad y_1^\prime = \frac{f_{n,m+1}}{f_{n,m}}, \quad y_2^\prime = \frac{h_{n,m+1}}{h_{n,m}}\,,
\end{eqnarray*}
\item Yang-Baxter map
$$ x_1^\prime\,=\,y_1\,P,\quad y_1^\prime\,=\,x_1\,\frac{1}{P},\quad x_2^\prime\,=\,y_2\,Q,\quad y_2^\prime\,=\,x_2\,\frac{1}{Q}, $$
where
\begin{eqnarray*}
&& P:= \frac{\alpha^3 (x_1-1) (x_2-1) + \beta^3 (y_1-1) (x_1 (x_2-1) + x_2 (y_2-1))}{\alpha^3 (x_1-1) (x_2-1) y_1 + \beta^3  (y_1-1) (x_2 y_2-1)} \,,\\
&& Q := \frac{\alpha^3 (x_2-1) (x_1 y_1-1) + \beta^3 x_2 (y_1-1) (y_2-1)}{\alpha^3 (x_2-1) (y_1 (x_1-1) + y_2 (y_1-1)) + \beta^3  (y_1-1) (y_2-1)} .
\end{eqnarray*}
\item Lax matrix for the Yang-Baxter map
$$M(x,y;\alpha) \,=\,\frac{1}{(\alpha^3-\lambda^3)^{1/3}} 
\left( \begin{array}{ccc} 
\alpha & 0 & \alpha \lambda (1-x) \\ 
\frac{\alpha \lambda x (1-y)}{y} & \frac{\alpha x}{y} & 0 \\ 
0 & \frac{-\lambda x}{\alpha^2 (1-x) (1-y)} & \alpha x \end{array} \right) $$
\end{enumerate}

\item Scaling invariance of system (\ref{sbsq-sys-2})

\begin{enumerate}
\item Invariants
\begin{eqnarray*}
x_1 = \frac{g_{n+1,m}}{g_{n,m}},\quad x_2 = \frac{r_{n+1,m}}{r_{n,m}},\quad y_1 = \frac{g_{n+1,m+1}}{g_{n+1,m}}, \quad y_2 = \frac{r_{n+1,m+1}}{r_{n+1,m}}\,,\\
x_1^\prime = \frac{g_{n+1,m+1}}{g_{n,m+1}},\quad x_2^\prime = \frac{r_{n+1,m+1}}{r_{n,m+1}},\quad y_1^\prime = \frac{g_{n,m+1}}{g_{n,m}}, \quad y_2^\prime = \frac{r_{n,m+1}}{r_{n,m}}\,,
\end{eqnarray*}
\item Yang-Baxter map
$$ x_1^\prime\,=\,y_1\,R,\quad y_1^\prime\,=\,x_1\,\frac{1}{R},\quad x_2^\prime\,=\,y_2\,S,\quad y_2^\prime\,=\,x_2\,\frac{1}{S}, $$
where
\begin{eqnarray*}
&& R:= \frac{\alpha^3 (x_1-1) (x_2(y_2-1) + y_1 (x_2-1)) + \beta^3 x_1 x_2 (y_1-1) (y_2-1)}{\alpha^3 (x_1-1) (x_2 y_2-1) y_1 + \beta^3 x_2 (y_1-1) (y_2-1)} \,,\\
&& S := \frac{\alpha^3 (x_1-1) (x_2-1) y_1 + \beta^3 x_2 y_2 (x_1 y_1-1) (y_2-1)}{\alpha^3 (x_1-1) (x_2-1) y_1 y_2 + \beta^3 (y_2-1) (x_2 (y_1-1)+y_1 (x_1-1))} .
\end{eqnarray*}
\item Lax matrix for the Yang-Baxter map
$$M(x,y;\alpha) \,=\,\frac{1}{(\alpha^3-\lambda^3)^{1/3}} 
\left( \begin{array}{ccc} 
\frac{\alpha x}{y} & 0 & \alpha \lambda \frac{1-y}{y} \\ 
\frac{\alpha \lambda (1-x)}{y} & \frac{\alpha}{y} & 0 \\ 
0 & \frac{-\lambda x}{\alpha^2 (1-x) (1-y)} & \alpha \end{array} \right) $$
\end{enumerate}

\end{itemize}

\section*{Acknowledgments}

PX was supported by the {\emph{Newton International Fellowship}} grant NF082473 entitled ``Symmetries and integrability of lattice equations and related partial differential equations'', which is run by The British Academy, The Royal Academy of Engineering and The Royal Society. FWN is supported by a Royal Society/Leverhulme Trust Senior Research Fellowship (2011-12).


\begin{thebibliography}{99}

\bibitem{ABS} Adler V. E., Bobenko A. I. and Suris Yu. B. (2003) Classification of integrable equations on quad-graphs. The consistency approach {\em Comm. Math. Phys.} {\bf{233}} 513--543

\bibitem{ABS-YB} Adler V. E., Bobenko A. I. and Suris Yu. B. (2004) Geometry of Yang--Baxter Maps: pencils of conics
and quadrirational mappings {\em Comm. Anal. Geom.} {\bf{12}} 967--1007

\bibitem{A} Atkinson J. (2008) B{\"a}cklund transformations for integrable lattice equations {\em{J. Phys. A: Math. Theor.}} {\bf{41}} 135202 (8pp)

\bibitem{BS} Bobenko A. I. and Suris Yu. B. (2002) Integrable systems on quad-graphs {\em{Int. Math. Res. Notices}} {\bf{2002}} 573--611

\bibitem{Dr} Drinfeld V.G. (1992) On some unsolved problems in quantum group theory, in: {\em{Quantum Groups, Lecture Notes in Mathematics}} Vol. 1510, Springer

\bibitem{H} Hietarinta J. (2011) Boussinesq-like multi-component lattice equations and multi-dimensional consistency 
{\em{J. Phys. A: Math. Theor.}} {\bf 44}, no. 16, 165204 (22 pp)

\bibitem{N1} Nijhoff F.W. (1996) On some ``Schwarzian Equations'' and their Discrete Analogues. in: Eds. A.S. Fokas and I.M. Gel'fand,
{\it Algebraic Aspects of Integrable Systems: In memory of Irene Dorfman}, (Birkh\"{a}user Verlag), 237--260

\bibitem{N}  Nijhoff F. W. (2002) Lax pair for the Adler (lattice Krichever-Novikov) system {\em Phys. Lett. A} {\bf{297}} 49--58

\bibitem{NPCQ} Nijhoff F. W., Papageorgiou V. G., Capel H. W. and Quispel G. R. W. (1992) The lattice Gel'fand-Dikii hierarchy {\emph{Inverse Problems}} {\bf{8}} 597--651

\bibitem{PTV} Papageorgiou V., Tongas A. and Veselov A. (2006) Yang-Baxter maps and symmetries of integrable equations on quad-graphs {\em{J. Math. Phys.}} {\bf{47}} 083502 (16pp)

\bibitem{SV} Suris Yu. B. and Veselov A. (2003) Lax matrices for Yang-Baxter maps {\em{J. Nonlin. Math. Phys.}} {\bf{10}} Suppl. 2, 223--230

\bibitem{V} Veselov A. P. (2003) Yang-Baxter maps and integrable dynamics {\em{Phys. Lett. A}} {\bf{314}} 214--221

\bibitem{W} Walker A. J. (2001) {\emph{Similarity reductions and integrable lattice equations}} Ph.D. thesis, University of Leeds

\bibitem{X} Xenitidis P. (2011) Symmetries and conservation laws of the ABS equations and corresponding differential-difference equations of Volterra type {\em{J. Phys. A: Math. Theor.}} {\bf{44}} 435201 (22pp)

\bibitem{XN} Xenitidis P. and Nijhoff F. (2011) Symmetries and conservation laws of lattice Boussinesq equations {\em{submitted to Phys. Lett. A}}

\end{thebibliography}
\end{document}